\documentstyle[prd,aps,eqsecnum,preprint,tighten,floats]{revtex}
\begin{document}
\draft

\preprint{\vbox{\hfill SMUHEP/96--06 \\
          \vbox{\hfill OHSTPY--HEP--T--96--034}
          \vbox{\vskip0.5in}
          }}

\title{Vacuum Structure of Two-Dimensional Gauge Theories\\ on
the Light Front}

\author{Gary McCartor}
\address{Department of Physics, Southern Methodist University, Dallas,
TX 75275}

\author{David G. Robertson and Stephen S. Pinsky}
\address{Department of Physics, The Ohio State University, Columbus,
OH 43210}

\date{\today}

\maketitle

\begin{abstract}
We discuss the problem of vacuum structure in light-front field theory
in the context of (1+1)-dimensional gauge theories.  We begin by
reviewing the known light-front solution of the Schwinger model,
highlighting the issues that are relevant for reproducing the
$\theta$-structure of the vacuum.  The most important of these are the
need to introduce degrees of freedom initialized on two different null
planes, the proper incorporation of gauge field zero modes when
periodicity conditions are used to regulate the infrared, and the
importance of carefully regulating singular operator products in a
gauge-invariant way.  We then consider SU(2) Yang-Mills theory in 1+1
dimensions coupled to massless adjoint fermions.  With all fields in
the adjoint representation the gauge group is actually SU(2)$/Z_2$,
which possesses nontrivial topology.  In particular, there are two
topological sectors and the physical vacuum state has a structure
analogous to a $\theta$ vacuum.  We formulate the model using
periodicity conditions in $x^\pm$ for infrared regulation, and
consider a solution in which the gauge field zero mode is treated as a
constrained operator.  We obtain the expected $Z_2$ vacuum structure,
and verify that the discrete vacuum angle which enters has no effect
on the spectrum of the theory.  We then calculate the chiral
condensate, which is sensitive to the vacuum structure.  The result is
nonzero, but inversely proportional to the periodicity length, a
situation which is familiar from the Schwinger model.  The origin of
this behavior is discussed.

\end{abstract}


\section{Introduction}

Light-front quantization \cite{dir49} has recently emerged as a
potentially powerful tool for the nonperturbative study of quantum
field theories \cite{reviews}.  The main advantage of this approach is
the apparent simplicity of the vacuum state, which leads to major
simplifications in the solution of the Hamiltonian eigenvalue problem.
Indeed, naive arguments suggest that the physical vacuum is trivial on
the light front.  In many theories of interest, however, the structure
of the vacuum plays an important physical role, giving rise to, e.g.,
spontaneous symmetry breaking, confinement, vacuum angles, etc.  It is
therefore necessary to understand how these phenomena can occur in
light-front field theory.

These issues have recently been discussed in a variety of contexts.
If one regulates the infrared by imposing periodic or antiperiodic
boundary conditions on some finite interval in $x^-$ \cite{pab85},
then any nontrivial vacuum structure must be connected with the
$k^+=0$ Fourier modes of the fields.\footnote{This follows from simple
kinematical considerations.}  Studies of model field theories have
shown that certain aspects of vacuum physics can in fact be reproduced
by a careful treatment of the field zero modes in this framework.  For
example, it has been shown that solutions of the zero mode constraint
equation in $\phi^4_{1+1}$ theory \cite{may76} exhibit spontaneous
symmetry breaking \cite{hek92a,rob93,bep93,piv94,hip95}.  In addition,
certain topological features of pure Yang-Mills theories in 1+1
dimensions have been successfully reproduced \cite{kap94}.

The focus of the present work is on structure of the $\theta$-vacuum
type in gauge theories.  This has been discussed in detail for the
Schwinger model in Refs. \cite{mcc91} (see also \cite{kar96} for a
discussion in the bosonized context), where it was shown that in order
to obtain a theory that is isomorphic to the usual equal-time theory
it is necessary to go beyond the conventional light-front approach.
The main complication is the need to introduce degrees of freedom
initialized along a second null plane, specifically a surface of
constant $x^-$.  In addition, it is important to properly treat the
gauge field zero modes and to carefully define singular operator
products in a gauge-invariant way.

Non-Abelian realizations of this sort of vacuum structure are
difficult to find in 1+1 dimensions, however, due to the fact that
$\Pi_1[{\rm SU}(N)]$ is trivial.  A model which does exhibit a sort of
$\theta$ vacuum is Yang-Mills theory coupled to fermions in the
adjoint representation.  Since all fields transform according to the
adjoint representation, gauge transformations that differ by an
element in the center of the gauge group represent the same
transformation and so should be identified.  Thus the gauge group is
actually ${\rm SU}(N)/Z_N$, which has nontrivial topology: $\Pi_1[{\rm
SU}(N)/Z_N]=Z_N$.  The model therefore possesses an $N$-fold vacuum
degeneracy, and there is a discrete vacuum angle analogous to the
$\theta$ parameter of QCD \cite{wit79,pas96}.  In addition, for $N=2$
there is expected to be a nonvanishing bilinear condensate
\cite{smi94}.

The goal of the present work is to understand how this structure
arises in the light-front framework.  As we shall see, if proper
attention is paid to the subtleties of light-front quantization, then
the expected features can all be correctly reproduced.  In particular,
for $N=2$ we shall explicitly exhibit the $Z_2$ vacuum degeneracy and
find a nonzero condensate.  In the light-front representation the
vacuum states can be described completely, unlike in the equal-time
approach.  However, the condensate we obtain is proportional to $1/L$,
where $L$ is the periodicity length, and so vanishes in the
infinite-volume limit.  This behavior is familiar from the Schwinger
model and may be traced to the infrared regulator we employ.  We shall
discuss these issues further below.

Interestingly, for massless fermions the spectrum of {\em massive}
states of the adjoint model has recently been shown to be identical to
that of YM$_{1+1}$ with multiple flavors of fundamental fermions
\cite{kus95}.  For this to be true it is necessary that the massive
spectrum be independent of the vacuum angle that enters in the
construction of the physical ground state.  We will show this
explicitly.  In fact, the only quantity which depends on the vacuum
angle is the chiral condensate, much like in the Schwinger model.

We shall begin by reviewing the essentials of the light-front solution
of the Schwinger model presented in Refs. \cite{mcc91}.  This will
serve to introduce the basic framework and to highlight the issues
that are central to the occurrence of nontrivial vacuum structure in
the light-front representation.  We then discuss the formulation of
SU(2) gauge theory with adjoint fermions.  We shall consider a
formulation of the theory in which the gauge field zero mode is
treated as a constrained variable; a complementary formalism, in which
the vacuum contains a dynamical zero mode content, is discussed in
Ref. \cite{pinsky96}.  We show that this model possesses degenerate
vacuum states which we calculate explicitly.  The physical ground
state is a superposition of these constructed to satisfy the cluster
property.  Next we compute the condensate, the expectation value of
$\bar{\Psi}\Psi$ in this state, and briefly discuss the $L$-dependence
of the result.  Finally, we touch on some unresolved issues and
directions for future work.

\section{The Schwinger Model}

The Schwinger model is electrodynamics of massless fermions in 1+1
dimensions \cite{sch62}.  The present discussion will necessarily be
rather telegraphic, as our aim is mainly to highlight the issues that
will be important later.  For further details the reader is advised to
consult Refs. \cite{mcc91}.

To begin with, let us consider a free massless fermion.  We shall
employ the convention $x^{\pm} = (x^0 \pm x^1)/ \sqrt 2$, and
decompose the Fermi field in the usual way:
\begin{equation}
\psi _\pm \equiv{1\over\sqrt{2}}\gamma^0\gamma^\pm\psi\; .
\end{equation}
In 1+1 dimensions (only) these are the same as chiral projections, so
that
\begin{equation}
\psi_+ = {\psi _R \choose 0}\; ,\qquad
\psi_- = {0 \choose \psi _L}\; .
\label{project}
\end{equation}
Now, the need to include degrees of freedom on two different
light-like lines can be seen immediately from the equation of motion,
which takes the form
\begin{equation}
\partial_+\psi_R = \partial_-\psi_L= 0\; ,
\end{equation}
where $\partial_\pm\equiv\partial/\partial x^\pm$.  These have as
their general solution
\begin{equation}
\psi_R= f(x^-)\; ,\qquad \psi_L= g(x^+)\; ,
\end{equation}
where $f$ and $g$ are arbitrary.  Clearly, information along lines of
both constant $x^+$ and $x^-$ must be specified to obtain a general
solution to the Dirac equation.  If in the quantum theory we do not
include degrees of freedom to represent all of this freedom of the
classical solution space, then the resulting theory will be
incomplete.

The proper light-front formulation of this theory involves a pair of
independent fields: $\psi_R$, initialized on $x^+=0$, and $\psi_L$,
initialized on $x^-=0$.  We regulate the infrared behavior by
requiring that $\psi_{R/L}$ satisfy antiperiodic boundary conditions
on $0 \leq x^\mp \leq 2L$.\footnote{Note that in general the
initial-value surface should be chosen so as to contain no points that
are separated by time-like intervals; in such a case the commutation
relations of the fields could not in general be known {\em a priori}.
For a detailed discussion of these issues in the light-front context
see \cite{rob96}.}  We can then Fourier expand the fields on their
respective initial-value surfaces:
\begin{eqnarray}
\psi_R(0,x^-) &=& {1\over2^{1/4}\sqrt{2L}}
\sum_{n={1\over2}}^\infty 
\left(b_n e^{-ik_n^+ x^-} + d^\dagger_n e^{ik_n^+ x^-}\right)\\
\psi_L(x^+,0) &=& {1\over2^{1/4}\sqrt{2L}}
\sum_{n={1\over2}}^\infty \left(\beta_n e^{-ik_n^- x^+}
+\delta^\dagger_n e^{ik_n^-x^+}\right)\; ,
\end{eqnarray}
where $k_n^\pm = n\pi/L$ and the sums are over odd half integers.
Throughout this paper we shall use lower-case (upper-case) letters to
denote indices that take odd half-integer (integer) values.  The
canonical anti-commutation relations are
\begin{eqnarray}
\left\{\psi_R(0,x^-),\psi_R^\dagger(0,y^-)\right\} & = &
{1\over\sqrt{2}}\delta(x^--y^-)
\label{smccr1}\\
\left\{\psi_L(x^+,0),\psi_L^\dagger(y^+,0)\right\} & = &
{1\over\sqrt{2}}\delta(x^+-y^+)\; .
\label{smccr2}
\end{eqnarray}
These are realized by the Fock algebra
\begin{equation}
\{b_n,b^\dagger_m\}=\{d_n,d^\dagger_m\}=
\{\beta_n,\beta^\dagger_m\}=\{\delta_n,\delta^\dagger_m\}
=\delta_{m,n}\; ,
\end{equation}
with all other anti-commutators vanishing.  The Fock space is
generated by applying the various creation operators to a vacuum state
$|0\rangle$.

An important feature of this construction is that the dynamical
operators $P^\pm$, and in fact all conserved charges, receive
contributions from both parts of the initial-value surface.  This
follows from very general considerations \cite{mcc88}.  We have
\begin{equation}
P^\pm=\int_0^{2L} dx^- \Theta^{+\pm}+\int_0^{2L} dx^+ \Theta^{-\pm}
\; ,
\label{smppm}
\end{equation}
where the second term accounts for the energy-momentum of the
left-movers.

Let us now turn to the Schwinger model.  The classical Lagrangian
density is
\begin{equation}
{\cal L} = \frac{1} {2} \left(i \bar{\psi} \gamma^{\mu}
\partial_{\mu}\psi-i(\partial_{\mu}\bar{\psi})\gamma^{\mu}\psi\right)
- \frac{1}{4} F^{\mu \nu} F_{\mu \nu} - eA^{\mu} J_{\mu} \; ,
\label{smlag}
\end{equation}
where $F_{\mu\nu} = \partial_\mu A_\nu - \partial_\nu A_\mu$ and
$J^\mu=\overline{\psi}\gamma^\mu\psi$.  We shall impose periodic
boundary conditions in $x^-$ on $A_\mu$, and choose the gauge
$\partial_-A^+=0$.  Note that the light-cone gauge $A^+=0$ is not
allowed in the presence of the nontrivial spatial topology
\cite{frn81}.  Thus $A^+ \equiv v$, a zero mode.  To simplify the
notation let us further introduce $A \equiv A^-$.  The equations of
motion then take the form
\begin{eqnarray}
-\partial_-^2A = eJ^+ &\equiv& eJ^R
\label{gauss}\\
\nonumber\\
\partial_+\partial_-A = eJ^- &\equiv& eJ^L
\label{ampere}\\
\nonumber\\
\left(\partial_+ + ieA\right)\psi_R &=& 0
\label{diracR}\\
\nonumber\\
\left(\partial_- + iev\right)\psi_L &=& 0\; ,
\label{diracL}
\end{eqnarray}
where $J^R=\sqrt{2}\psi_R^\dagger\psi_R$ and $J^L =
\sqrt{2}\psi_L^\dagger\psi_L$.

We should perhaps elaborate somewhat on the choice of gauge.  The
gauge field is chosen to be periodic in $x^-$ but satisfies no
particular boundary condition in $x^+$.  Thus the gauge transformation
required to bring an arbitrary configuration $A_\mu(x^+,x^-)$ to one
satisfying $\partial_-A^+=0$ will in general not be periodic in $x^+$,
and so will violate the boundary condition we have imposed on
$\psi_L$.  However, after such a transformation we can apply a purely
$x^+$-dependent gauge transformation that restores the antiperiodicity
of $\psi_L$ at a single value of $x^-$, which we can choose to be the
initial-value surface $x^-=0$.  This does not affect the boundary
conditions satisfied by the other fields.  To be precise, therefore,
we should say that we require that $\psi_L$ be antiperiodic on its
initial-value surface only; it may not remain antiperiodic as it
evolves in $x^-$.  (Exactly what happens to $\psi_L$ is discovered by
solving its equation of motion.)  This condition, and the conditions
imposed on $\psi_R$ and $A_\mu$, are then consistent with the gauge
choice $\partial_-A^+=0$.

Next let us discuss the definition of singular operator products, as
this is central to the issue of vacuum structure.  We define the
current operators by a gauge-invariant point splitting:
\begin{equation}
J^R(0,x^-) \equiv\sqrt{2}\lim_{\epsilon^-\rightarrow0}
\left(e^{-ie\int_x^{x+\epsilon^-} v dx^-}
\psi_R^\dagger(0,x^-+\epsilon^-) \psi_R(0,x^-)-{\rm VEV}\right)
\label{jplusdef}
\end{equation}
\begin{equation}
J^L(x^+,0) \equiv\sqrt{2}\lim_{\epsilon^+\rightarrow0}
\left(e^{-ie\int_x^{x+\epsilon^+} A dx^+}
\psi_L^\dagger(x^++\epsilon^+,0) \psi_L(x^+,0)-{\rm VEV}\right)\; .
\label{jminusdef}
\end{equation}
Note that we must split $\psi_R^\dagger\psi_R$ in the $x^-$ direction
and $\psi_L^\dagger\psi_L$ in the $x^+$ direction.  This follows from
the canonical singularity structure of the fields [Eqns. (\ref{smccr1})
and (\ref{smccr2})].  Evaluating the singularities in the operator
products as $\epsilon^\pm \rightarrow0$ we find
\begin{eqnarray}
J^R & = & \tilde{J}^R -{e\over2\pi}v \\
J^L & = & \tilde{J}^L -{e\over2\pi}A\; ,
\end{eqnarray}
where $\tilde{J}^\pm$ are the ``naive'' normal-ordered currents.  It
will be useful to express these in terms of their Fourier modes (the
so-called ``fusion operators'').  We write
\begin{eqnarray}
\tilde{J}^R  & = & {1\over2L}\sum_{N = -\infty}^\infty C_N e^{-ik_N^+x^-}
\label{thecs}\\
\tilde{J}^L & = & {1\over2L}\sum_{N = -\infty}^\infty D_N
e^{-ik_N^-x^+}\; ,
\label{theds}
\end{eqnarray}
where the sums run over the integers.  For $N=0$ these are the charge
operators for the right- and left-movers,
\begin{eqnarray}
C_0 & = & \sum_{n={1\over2}}^\infty
\left(b^\dagger_nb_n - d^\dagger_nd_n\right)\\ D_0 & =
& \sum_{n={1\over2}}^\infty \left(\beta^\dagger_n\beta_n -
\delta^\dagger_n\delta_n\right)\; ,
\end{eqnarray}
while for $N>0$ they are given by
\begin{eqnarray}
C_N & = & \sum_{n={1\over2}}^\infty 
\left(b^\dagger_nb_{N+n} - d^\dagger_nd_{N+n}\right)
+ \sum_{n={1\over2}}^{N-{1\over2}}d_{N-n}b_n \\
D_N & = & \sum_{n={1\over2}}^\infty \left(\beta^\dagger_n\beta_{N+n}
-\delta^\dagger_n\delta_{N+n}\right) +
\sum_{n={1\over2}}^{N-{1\over2}} \delta_{N-n}\beta_n\; .
\end{eqnarray}
For $N<0$ they may be obtained by conjugation:
\begin{equation}
C_{-N} = C_N^\dagger\qquad ,\qquad D_{-N} = D_N^\dagger\; .
\end{equation}
They can be shown to satisfy the simple algebra
\begin{equation}
\left[C_M,C_N\right] = \left[D_M,D_N\right] = M\delta_{M,-N}\; .
\label{simplealg}
\end{equation}

We can now discuss the implementation of Gauss' law,
Eqn. (\ref{gauss}).  Projected onto the normal-mode sector (in $x^-$),
this is a constraint which determines the normal modes of $A$ on
$x^+=0$ in the usual way.  Projected onto the zero-mode sector we
obtain
\begin{equation}
0 = C_0 - z_R \; .
\label{gausszm}
\end{equation}
where $z_R \equiv e v L/ \pi$.  This can be interpreted in at least
two ways.  One obvious possibility is to just use Eqn. (\ref{gausszm})
to determine $z_R$ in terms of the charge operator (although there is
a subtlety that arises which we shall explain in a moment).  If we
wish to treat $z_R$ as a dynamical field, however, then
Eqn. (\ref{gausszm}) cannot hold as an operator relation; it must be
implemented via restriction to some physical subspace.  The principal
difficulty with this approach arises from the fact that $z_R$ is
$x^+$-dependent while $C_0$ is not.  Thus the operator we would like
to use to select physical states does not commute with the Hamiltonian
$P^-$, and the stability of the physical subspace is unclear.  It is
immediately obvious, for example, that the vanishing of the RHS of
(\ref{gausszm}) cannot be imposed as an eigenvalue condition, if there
are to be eigenstates of the Hamiltonian in the physical subspace.  It
may be possible to realize this condition in matrix elements between
suitably defined physical states, however.

Here we shall interpret Eqn. (\ref{gausszm}) as determining $z_R$ in
terms of the charge.  There is a subtlety in solving (\ref{gausszm})
as it stands, however.  We shall see below that $z_R$ should not be
taken to be a function of $\psi_R$, as this would lead to
inconsistencies in the Heisenberg equations.  Anticipating that the
physical subspace of the Schwinger model will consist of states with
vanishing {\em total} charge $C_0+D_0$, let us rewrite
Eqn. (\ref{gausszm}) in the form
\begin{equation}
0 = C_0 + D_0-D_0-z_R\; .
\end{equation}
We can then take
\begin{equation}
z_R= - D_0\; ,
\label{apzm}
\end{equation}
which gives no conflict with the Heisenberg equations, and what
remains is the expected condition defining physical states.  Thus the
zero mode of Gauss' law is interpreted as determining $z_R$ in terms
of $D_0$, and giving the condition $C_0+D_0=0$, which must be imposed
as an eigenvalue condition on the states.  This construction may seem
somewhat {\em ad hoc} but in fact it can be justified by careful
consideration of the proper coupling of the gauge field to matter
\cite{mcc91}.  The situation is really the same as in classical
electromagnetism.

The zero mode of Ampere's law, Eqn. (\ref{ampere}), may be analyzed
similarly, leading to
\begin{equation}
z_L= - C_0\; ,
\label{amzm}
\end{equation}
where $z_L \equiv e w L / \pi$ and $w$ is the zero mode of $A$, and
the same neutrality condition on the states, $C_0+D_0=0$.

Next let us construct the Poincar\'e generators $P^\pm$.  As discussed
above, these are given as integrals of the appropriate components of
the energy-momentum tensor over both pieces of the initial-value
surface.  If we work out $\Theta^{\mu\nu}$ as usual and try to
evaluate (\ref{smppm}), however, we encounter a difficulty: the
integral over $x^+$ involves fields that are initialized on the
surface $x^+=0$.  But we do not know these fields as functions of
$x^+$ until we have solved the theory!  There is a simple way of
dealing with this which works, at least for the Schwinger model.  We
work out $\Theta^{\mu\nu}$ as usual, but include in the calculation of
$P^\pm$ only those terms that contain quantities we know on the
different parts of the initial-value surface.

No proof exists of the correctness of this procedure, but any alleged
results for $P^\pm$ can be checked {\em a posteriori} for correctness,
by verifying that they properly translate all the fields in $x^\pm$
and satisfy the Poincar\'e algebra.  The same is true of other
operators, for example the charges or the boost generator---we can
check in the end whether the expressions we take are the correct ones.
We will approach the problem of constructing $P^\pm$ from this
practical perspective: we start from the canonical energy-momentum
tensor, but modify it as necessary in response to any problems of
consistency which arise.  In the end we shall justify the final
expressions by showing that they satisfy all necessary criteria.

Let us first consider $P^+$.  The relevant components of the
energy-momentum tensor, derived via the standard Noether procedure,
are
\begin{eqnarray}
\Theta^{++} &=& i\sqrt{2}\psi_R^\dagger\partial_-\psi_R\\
\nonumber\\
\Theta^{-+} &=& -i\sqrt{2}\psi_R^\dagger\partial_+\psi_R
+e\left(J^R A+J^Lv\right)\; .
\end{eqnarray}
Using the rule stated above the only contribution comes from
$\Theta^{++}$, so that
\begin{equation}
P^+ = i\sqrt{2}\int_0^{2L} dx^- \psi_R^\dagger\partial_-\psi_R\; .
\end{equation}
The operator product that occurs is singular, however, and must be
regulated.  We shall again split the product in $x^-$, introducing an
eikonal factor to maintain gauge invariance:
\begin{equation}
\psi_R^\dagger\partial_-\psi_R \equiv \lim_{\epsilon^-\rightarrow 0}
\left( e^{-ie\int_{x^-}^{x^-+\epsilon^-} dy^-v}
\psi_R^\dagger(x^-+\epsilon^-) \partial_-\psi_R(x^-) - {\rm
VEV}\right)\; .
\end{equation}
(Note that in some cases a symmetric splitting may be necessary to
maintain the hermiticity of the regulated operator.)  Evaluating the
singularity in $\psi_R^\dagger\partial_-\psi_R$ as $\epsilon^-
\rightarrow 0$ we then obtain
\begin{equation}
P^+ = P^+_{\rm free} - {\pi\over 2L}z_R^2\; ,
\label{p+wgc}
\end{equation}
where $P^+_{\rm free}$ is the free-particle momentum operator for the
right-movers:
\begin{equation}
P^+_{\rm free} = \sum_n \left( {n\pi\over L}\right) \left[b^\dagger_n
b_n + d^\dagger_n d_n\right]\; .
\end{equation}
It is now possible to see why it would be dangerous to have $z_R$ be a
function of the right-movers, through $C_0$, for example.  The
operator $P^+$ should generate translations of $\psi_R$ in its
initial-value surface via
\begin{equation}
[\psi_R,P^+] = i\partial_-\psi_R\; .
\end{equation}
But this is already accomplished by $P^+_{\rm free}$, and any
additional terms in $P^+$ should commute with $\psi_R$ to avoid
spoiling this relation.  This is what leads us to the modification of
the zero mode part of Gauss' law discussed above, and to the solution
(\ref{apzm}) for $z_R$ in terms of $D_0$.

In fact, Eqn. (\ref{apzm}) is essentially forced on us by considering
the Heisenberg equation corresponding to Eqn. (\ref{diracL}).  We must
have
\begin{equation}
[\psi_L,P^+] = {1\over2L}\left(z_R\psi_L + \psi_L z_R\right)\; ,
\label{pmheis}
\end{equation}
or equivalently
\begin{equation}
[\psi_L,z_R] = -\psi_L\; .
\end{equation}
This requirement, along with $[\psi_R,z_R]=0$, essentially fixes $z_R$
to be given by Eqn. (\ref{apzm}).  Thus $P^+$ takes the final form
\begin{equation}
P^+ = P^+_{\rm free} -{\pi\over 2L} D_0^2\; .
\label{fullp+}
\end{equation}
Note the minus sign in front of the second term.  This is crucial for
the appearance of nontrivial vacuum structure, as it allows certain
states containing left-movers to be degenerate with the Fock vacuum
$|0\rangle$.

If we treat $z_R$ as dynamical, then Eqn. (\ref{p+wgc}) is actually
inconsistent as it stands.  This is because $\pi_R$, the momentum
conjugate to $z_R$, does not commute with this $P^+$.  But it {\em
must} commute with $P^+$, since it is itself a zero mode, i.e.,
independent of $x^-$, and $P^+$ generates translations on $x^-$.  This
problem may be solved by modifying $P^+$, but it is interesting to
note that a naive application of point-splitting appears to be
inconsistent with treating $z_R$ as dynamical.

The Hamiltonian $P^-$ can be constructed similarly.  The only subtlety
here involves the gauge correction of the singular operator product
occurring in
\begin{equation}
i\sqrt{2}\int_0^{2L} dx^+ \psi_L^\dagger \partial_+\psi_L\; .
\end{equation}
Because this product is split in the $x^+$ direction, the eikonal
factor involves $A$.  But most of $A$ is unknown on the surface
$x^-=0$; its normal modes are constrained functionals of $\psi_R$,
determined by solving Gauss' law.  The solution to this difficulty is
to keep in the expression for $P^-$ only that part of $A$ that we do
know on $x^-=0$, namely its zero mode (\ref{amzm}).  (That $C_0$ is
$x^+$-independent will be checked momentarily.)  With this Ansatz we
arrive at
\begin{equation}
P^- = P^-_{\rm free} - {\pi\over 2L} C_0^2 + {e^2L\over2\pi^2}
\sum_{N=-\infty}^{\infty} \!\!\!\!{}^{\stackrel{\prime}{}} \;
\left({1\over N^2}\right) C_N C_{-N}\; ,
\end{equation}
where the prime indicates that the term with $N=0$ is omitted and
$P^-_{\rm free}$ is the free energy of the left-movers:
\begin{equation}
P^-_{\rm free} = \sum_n \left({n\pi\over L}\right)
\left[\beta^\dagger_n \beta_n +\delta^\dagger_n \delta_n\right]\; .
\end{equation}
It is clear that this $P^-$ commutes with $C_0$, so that $\partial_+
C_0=0$ as promised.

At this stage we have dynamical operators that correctly translate all
the fermionic degrees of freedom in $x^\pm$, and are consistent with
the solutions we took for the zero modes of the gauge field.  Gauss'
law is also satisfied by construction.  The only issue that remains is
whether or not Ampere's law, Eqn. (\ref{ampere}), is satisfied.

This is straightforward to check.  The zero mode of Ampere's law
reduces to the condition $C_0+D_0=0$, which defines the physical
subspace.  To check the normal mode part, we take the operator
$\partial_-A$ obtained from solving Gauss' law and commute it with
$P^-$ to obtain its $x^+$-derivative.  It is straightforward to check
that this results in an equality, except that the terms in $J^L$
involving the $D_N$ are not reproduced in the commutator.  Thus
Ampere's law is only obtained in matrix elements between states that
satisfy
\begin{equation}
D_N|{\rm \Phi}\rangle = 0 \qquad (N>0)\; ,
\end{equation}
This must be added to the conditions defining physical states.

This is a very important result, as it removes most of the states with
left-moving quanta from the physical subspace.  In fact, it can be
shown \cite{mcc91} that the {\em only} states with left-movers that
remain in the physical subspace are the states
\begin{eqnarray}
|V_N\rangle &=& d^\dagger_{N-\frac{1} {2}}\beta^\dagger_{N-\frac{1}
{2}}\dots
      d^\dagger_\frac{1} {2} \beta^\dagger_\frac{1} {2} |0\rangle
\label{vsubn} \\
\nonumber\\
|V_{-N}\rangle &=& b^\dagger_{N-\frac{1}{2}}\delta^\dagger_{N-\frac{1}
{2}}\dots
      b^\dagger_\frac{1} {2} \delta^\dagger_\frac{1} {2} |0\rangle
\label{vsub-n}
\end{eqnarray}
where $N=1,2,\dots$, and states created by applying functionals of
$\psi_R$ to these.  It is straightforward to check that these are all
eigenstates of $P^\pm$ with eigenvalue zero, so that they are
degenerate with the bare vacuum.  The physical ground state of the
theory is a superposition of these, which can be shown to be necessary
to satisfy the requirement of cluster decomposition:
\begin{equation}
|\theta\rangle = \sum_{N=-\infty}^\infty e^{-iN\theta}|V_N\rangle\; .
\end{equation}
Thus we recover the expected $\theta$-structure of the physical vacuum
state in this model.

The full solution of the theory is straightforward given the form of
$P^-$.  We first verify that the states obtained by acting with the
$C_N$ on $|\theta\rangle$ span the physical subspace.  We then define
creation and destruction operators from the positive and negative
frequency $C_N$ respectively.  Eqn. (\ref{simplealg}) is then a
bosonic canonical commutator and the Hamiltonian is diagonal.  In the
end we recover the well-known results that the physical states
correspond to noninteracting bosons of mass $e/\sqrt{\pi}$.  In
addition, there is a nonvanishing condensate $\langle\theta|
\bar{\psi}\psi |\theta\rangle$ with the correct dependence on
$\theta$.  One can also check that the correct chiral anomaly is
obtained.  These and other issues are discussed in detail in
Refs. \cite{mcc91}.

Our goal in this section was to use the Schwinger model to highlight
the issues that will be important when we discuss YM$_{1+1}$.  The
main lessons to be drawn concern the need to include degrees of
freedom initialized on two different null planes, and the regulation
of singular operator products in a gauge-invariant way.  In addition,
we must construct dynamical operators that are consistent with the
equations of motion and initial conditions, and identify a subspace in
which those equations of motion that are not satisfied as operator
relations can hold.  Finally, we must verify that this physical
subspace is stable under evolution in $x^\pm$.

\section{SU(2) Gauge Theory Coupled to Adjoint Fermions: Basics}

Let us now consider SU(2) gauge theory coupled to massless adjoint
fermions in 1+1 dimensions.  The Lagrangian density for the theory is
\begin{equation}
{\cal L} = - {1 \over 2} Tr (F^{\mu \nu} F_{\mu \nu}) + {i \over 2}
Tr(\bar\psi\gamma ^{\mu} \buildrel \leftrightarrow \over
D_{\mu}\psi)\; ,
\end{equation}
where $D_{\mu} = \partial_{\mu} + ig [ A_\mu,\ \ ] $ and $F_{\mu \nu}
= \partial_{\mu}A_{\nu} - \partial_{\nu} A_{\mu} + ig [A_{\mu},
A_{\nu} ]$.  A convenient representation for the gamma matrices is
$\gamma ^0 = \sigma^2$ and $\gamma^1 = i\sigma^1$, where $\sigma^a$
are the Pauli matrices.  With this choice the (Majorana) Fermi field
may be taken to be hermitian.

The matrix representation of the fields makes use of the fundamental
SU(2) generators $\tau^a=\sigma^a/2$.  It is convenient to introduce a
color helicity, or Cartan, basis, defined by
\begin{equation}
\tau^\pm\equiv {1\over\sqrt{2}}\left(\tau^1\pm i\tau^2\right)\; ,
\end{equation}
with $\tau^3$ unchanged.  These satisfy
\begin{eqnarray}
\left[\tau^+, \tau^- \right] &=& \tau^3 \\
\left[\tau^3 , \tau^{\pm} \right] &=& \pm\tau^{\pm}\; .
\end{eqnarray}
Lower helicity indices are defined by $\tau_\pm = \tau^\mp$, and
matrix-valued fields are decomposed as, for example,
\begin{equation}
A^{\mu} = A _3 ^{\mu} \tau ^3 + A ^{\mu} _+ \tau^+ + A_-^{\mu}\tau^-
\; ,
\end{equation}
where $A^{\mu,\pm}\equiv (A^\mu_1\pm iA^\mu_2)/ \sqrt{2}$ and
$A^{\mu,\pm} = A^\mu_\mp$.  (Note also that
$(A^\mu_+)^\dagger=A^\mu_-$.)  The Fermi field will be written as
\begin{equation}
\Psi _{R/L} = \psi _{R/L} \tau ^3 + \phi _{R/L} \tau ^+ +
\phi^\dagger _{R/L} \tau ^-\; ,
\label{helicity}
\end{equation}
where $\phi_{R/L}\equiv (\Psi^1_{R/L}-i\Psi^2_{R/L})/\sqrt{2}$ and the
labels $R/L$ indicate light-front spinor projections as given in Eqn.
(\ref{project}).  Note that under a gauge transformation the Fermi
field transforms according to
\begin{equation}
\Psi _{R/L}^\prime = U \Psi _{R/L} U^{-1}\; ,
\end{equation}
where $U$ is a spacetime-dependent element of SU(2).

We shall regulate the theory in the infrared by imposing certain
boundary conditions in $x^\pm$.  The fields $\phi_R$ and $\phi_L$ will
be taken to be antiperiodic in $x^-$ and $x^+$, respectively.  It will
be convenient, however, to take $\psi_R$ and $\psi_L$ to be periodic
in $x^-$ and $x^+$, respectively (``twisted'' boundary conditions).
The reasons for this will become clear as we progress.  For
consistency, then, $A^\mu_\pm$ must be taken to be antiperiodic in
$x^-$, while $A^\mu_3$ is periodic.  In all cases the periodicity
length is $2L$.

The Fock representation for the fermionic degrees of freedom is
obtained by Fourier expanding $\Psi_R$ on $x^+=0$ and $\Psi_L$ on
$x^-=0$.  We have
\begin{eqnarray}
\psi _R(0,x^-) &=& {1 \over 2 ^{1/4} \sqrt {2L}} \sum_{N=1}^\infty 
\left(a_N e^{-ik {^+ _N} x^-} + a{_N ^\dagger} e^{ik{_N ^+}x^-} \right)
+{\stackrel {o} {\psi}}_R \\
\phi _R(0,x^-)&=& {1 \over 2 ^{1/4} \sqrt {2L}} \sum_{n={1\over2}}^\infty 
\left(b_n e^{-ik {^+_n} x^-} + d{_n ^\dagger} e^{ik{_n ^+}x^-} \right)
\label{phirinitial} \\
\psi _L(x^+,0) &=& {1 \over 2 ^{1/4} \sqrt {2L}} \sum_{N=1}^\infty
\left(\alpha_N e^{-ik{^- _N} x^+} + \alpha{_N ^\dagger} 
e^{ik{_N ^-}x^+} \right) +{\stackrel {o} {\psi}}_L \\
\phi _L(x^+,0) &=& {1 \over 2 ^{1/4} \sqrt {2L}} \sum_{n={1\over2}}^\infty
\left(\beta_n e^{-ik{^- _n} x^+} + \delta{_n ^\dagger} 
e^{ik{_n ^-}x^+} \right) \; ,
\end{eqnarray}
where we have explicitly separated out the zero modes of $\psi_{R/L}$.
As before, the lower-case (upper-case) indices run over positive
half-odd integers (integers) and $k{^\pm _n}= n \pi /L$.  The Fourier
modes obey the algebra
\begin{equation}
\{a{^\dagger _N}, a_M \} = \{\alpha{^\dagger _N}, \alpha_M \} 
= \delta_{N,M}
\label{rhccrs}
\end{equation}
\begin{equation}
\{ b{_n ^\dagger}, b_m \} = \{ d{_n^\dagger}, d_m \} 
= \{ \beta{_n ^\dagger},\beta _m \} 
= \{\delta {_n^\dagger},\delta_m\} = \delta _{n, m}
\label{lhccrs}
\end{equation}
\begin{equation}
\{{\stackrel {o} {\psi}}_R,{\stackrel {o} {\psi}}_R \} = 
\{{\stackrel {o} {\psi}}_L,{\stackrel {o} {\psi}}_L \} = 
{1\over2\sqrt{2} L}\; ,
\label{zmccrs}
\end{equation}
with all mixed anti-commutators vanishing.  These are equivalent to
the canonical anti-commutation relations
\begin{eqnarray}
\left\{\Psi_R(0,x^-),\Psi_R(0,y^-)\right\}
&=& {1\over\sqrt{2}} \delta (x^--y^-)\\
\left\{\Psi_L(x^+,0),\Psi_L(y^+,0)\right\}
&=& {1\over\sqrt{2}} \delta (x^+-y^+)\; .
\end{eqnarray}
The fermionic Fock space is generated by acting with the various
creation operators on a vacuum state $|0\rangle$.

For simplicity, in the remainder of this paper we shall discard the
zero modes of $\psi_{R/L}$.  It can be shown that including them does
not qualitatively affect any of our results; they merely complicate
parts of the analysis.  In addition, their physical meaning is
somewhat ambiguous.  For example, they lead to a nonvanishing fermion
condensate even in free field theory!  This same phenomenon was
observed in the equal-time context in Ref. \cite{nak80}.  We shall
therefore simply exclude them from the model; the condensate we obtain
is then entirely an effect of the interaction.  Note that this
truncation does not lead to inconsistencies in the model.  For
example, the Heisenberg equations for $\psi_{R/L}$ will simply reduce
to the appropriate Euler-Lagrange equations with the zero modes
removed.  In addition, the Poincar\'e algebra is unaffected.

The current operators for this theory are
\begin{eqnarray}
J^+ \equiv J^R &=& -{1\over\sqrt 2}[\Psi _{R},\Psi_R]
\label{jplus} \\
\nonumber\\
J^- \equiv J^L &=& -{1\over\sqrt 2}[\Psi _{L},\Psi_L]\; .
\label{jminus}
\end{eqnarray}
To avoid confusion, we shall henceforth always write the currents with
$R$ or $L$ in place of the upper Lorentz index.  These expressions are
ill-defined as they stand since they contain products of operators at
the same point. This is a common problem and occurs in the expressions
for the Poincar\'e generators as well.  We shall regulate these by
point splitting, introducing an eikonal factor to maintain gauge
invariance, and then taking the splitting to zero after removing the
divergences.  One can show for example that
\begin{equation}
\left[
e^{ ig \int_x^{x + \epsilon^\mu}  A \cdot dx}\Psi(x+\epsilon^\mu)
e^{-ig \int^{x +\epsilon^\mu}_x   A \cdot dx},\Psi(x)
\right]
\label{ham}
\end{equation}
transforms covariantly in the adjoint representation. In the limit of
small $\epsilon^\mu$ one effectively replaces
\begin{equation}
\Psi(x) \rightarrow \Psi(x+\epsilon^\mu) +ig [ A \cdot \epsilon, 
\Psi(x+\epsilon^\mu) ]\; .
\end{equation}
The singularity in the Fermi operator product as $\epsilon
\rightarrow0$ picks up the $\epsilon$ in the above expression leaving
an additional contribution.  The splitting must be performed in the
$x^-$ direction for $\Psi_R$ and in the $x^+$ direction for $\Psi_L$.
A straightforward calculation gives
\begin{eqnarray}
J^R &=& \tilde{J^R} - {g \over 2\pi} A^+
\label{Ranomaly}\\
J^L &=& \tilde{J^L} - {g \over 2\pi} A^- \; ,
\label{Lanomaly}
\end{eqnarray}
where $\tilde{J}^{R/L}$ are the naive normal-ordered currents.

In the helicity basis the components of $\tilde{J}^R$ take the forms
\begin{eqnarray}
\tilde{J}^R_3 &=& :{1\over\sqrt{2}} \left(
\phi^\dagger_{R} \phi_{R}-\phi_{R} \phi^\dagger_{R} \right): \\
\nonumber\\
\tilde{J}^R_- &=& :{1\over\sqrt{2}} \left(
\psi_{R} \phi^\dagger_{R}-\phi^\dagger_{R} \psi_{R} \right): \\
\nonumber\\
\tilde{J}^R_+ &=& :{1\over\sqrt{2}} \left(
\phi_{R} \psi_{R}-\psi_{R} \phi_{R} \right): \; .
\end{eqnarray}
The corresponding expressions for the components of $\tilde{J}^L$ are
identical, with $R\rightarrow L$.  It is convenient to Fourier expand
these currents and discuss the properties of their components.  We
write
\begin{eqnarray}
\tilde{J}^{R,a} &=& {1\over 2L}
\sum_{N=-\infty}^{\infty} C^a_N e^{-i\pi N x^-/L}\\
\tilde{J}^{L,a} &=& {1\over 2L}
\sum_{N=-\infty}^{\infty} D^a_N e^{-i\pi N x^+/L}
\; ,
\end{eqnarray}
where $a$ is a color index and the sums run over integers for $a=3$
and half-odd integers for $a=\pm$.  It is well known that the Fourier
components satisfy a Kac-Moody algebra with level two \cite{goo86}.
We shall discuss this explicitly for the $C_N^a$; with appropriate
substitutions an identical set of relations holds for the $D_N^a$.

In terms of the Fock operators we have, for $N,n>0$,
\begin{eqnarray}
C_N^3 &=& \sum_{n={1 \over 2}}^\infty \left(b^\dagger_n b_{N+n}
-d^\dagger_n d_{N+n}\right)
-\sum_{n={1 \over 2}}^{N-{1 \over 2}} b_n d_{N-n} \\
C_n ^+ &=& \sum_{M=1}^\infty a^\dagger_M d_{n+M}
-\sum_{m={1 \over 2}}^\infty b^\dagger_m a_{n+m}
-\sum_{m={1 \over 2}}^{n-1} d_m a_{n-m} \\
C_n ^- &=& \sum_{m={1 \over 2}}^\infty d^\dagger_m a_{n+m}
-\sum_{M=1}^\infty a^\dagger_M b_{M+n}
-\sum_{m={1 \over 2}} ^{n-1} a_{n-m} b_m \; .
\end{eqnarray}
We may obtain the modes for $N,n<0$ from the above by hermitian
conjugation:
\begin{equation}
C^3_{-N} = (C^3_N)^\dagger \qquad C^+_{-n} = (C^-_n)^\dagger
\qquad C^-_{-n} = (C^+_n)^\dagger\; .
\end{equation}
Finally, $C^3_0$ is just the 3 color charge in the right-moving
fermions:
\begin{equation}
C^3_0 = \sum_n (b _n ^\dagger b_n - d_n ^\dagger d_n ) \; .
\end{equation}
In the Cartan basis, the Kac-Moody algebra takes the form
\begin{eqnarray}
\left [ C^3_N , C^3_M         \right ] &=& N \delta_{N,-M} \\
\left [ C_n^{\pm} , C_m^{\pm} \right ] &=& 0 \\
\left [ C^3_N , C_m^{\pm}     \right ] &=& \pm C_{N+m}^{\pm}
\label{kacmoody3pm}\\
\left [ C^+_n , C^-_m         \right ] &=& C^3_{n+m} + n \delta_{n,-m}\; .
\end{eqnarray}
It is straightforward to verify these relations using the fundamental
anti-commutators (\ref{rhccrs}) and (\ref{lhccrs}).  The algebra
satisfied by the $D$s is of course identical.

\section{Zero Modes and Gauge Fixing}

The main subtlety arising from the use of discretization as a
regulator is in fixing the gauge.  It is most convenient in
light-front field theory to choose the light-cone gauge $A^+=0$.
However, this is not possible with the boundary conditions we have
imposed; since gauge transformations must be periodic up to an element
of the center of the gauge group (here $Z_2$), we cannot gauge the
zero mode of $A^+$ to zero \cite{frn81}.  It is permissible to take
$\partial_-A^+=0$, as in the Schwinger model.  Having made that
choice, there are two issues which arise involving the zero modes of
the gauge field.

The first issue is whether the zero modes should be treated as
independent degrees of freedom or as constrained functionals of the
Fermi fields.  In the equal-time representation this can usually be
resolved on the basis of whether or not the Lagrangian gives a
conjugate momentum for the operator in question.  In the light-front
representation the issue is more complex; in the Schwinger model, for
example, the zero modes of the gauge fields are constrained even
though the Lagrangian provides a conjugate momentum.

Some of the difficulties in treating the zero mode $A^+$ as a degree
of freedom were mentioned in Sect. II.  They arise principally because
of Eqn. (\ref{gausszm}), which cannot hold as an operator relation.
Imposing it as a condition on states is somewhat delicate, however,
since the stability of the physical subspace is not obvious.  These
and other aspects of the formulation with a dynamical zero mode are
discussed in Ref. \cite{pinsky96}.  Here we shall avoid this
difficulty by considering a formulation in which the zero mode is
constrained.

In addition, $A^+$ enters the operator $P^+$ if we compute the
Poincar\'e generators according to the usual prescription.  But if
$A^+$ is dynamical, then the resulting $P^+$ does not commute with the
momentum conjugate to $A^+$.  This difficulty can perhaps be resolved
by replacing the troublesome terms in $P^+$ with operators to which
they are weakly equivalent (i.e., equal in the physical subspace).
Note that we must apply this procedure in the constrained case as
well.  If in the Schwinger model we had taken $z_R$ to be a functional
of $\psi_R$ (possibly along with other things) we would have found an
immediate contradiction between our definitions and the Heisenberg
equations.  Such a solution would be inconsistent even prior to any
dynamical considerations.  That is why in the Schwinger model $z_R$ is
taken to be a functional of $\psi_L$ (although there are other ways to
reach the same conclusion).  The non-Abelian case turns out to be even
more complicated, as we shall see.

The second issue involving the zero modes is the question of how to
accomplish the residual gauge fixing using constant (in $x^+$) color
matrices.  In the equal-time representation, it is convenient to
rotate the zero modes of $A^+$ so that only the color 3 component is
nonzero.  Here we shall mostly be interested in the case where the
zero modes of the gauge field are treated as constrained.  In this
case, since we should fix the gauge in terms of the degrees of freedom
and let the constraint equations determine the constrained variables,
we should proceed in a different way.  The natural thing would
presumably be to use the residual gauge freedom to rotate two
components of the vector current to zero; probably that is the choice
most nearly equivalent to rotating the gauge field zero modes in the
equal-time representation.  The problem is that this approach is
technically difficult to implement.  The Dirac-Bergmann procedure
leads to a complicated nonlinear relation between the Fermi modes, due
to the fact that the current is bilocal in the Fermi fields.  A
procedure which is very similar to the suggested gauge fixing, but is
much simpler to carry out, is to use twisted boundary conditions for
the Fermi fields \cite{tho81,shs94} as we do here.  Presumably the
results of the Dirac-Bergmann procedure would be much the same if we
were able to carry it out, but we have no proof of this.  With these
boundary conditions, the color 1 and 2 components of the gauge field
must also be antiperiodic in $x^-$ and simply have no zero modes.
Thus whether or not we treat the zero modes of $A^+$ as constrained we
have
\begin{equation}
A^+ = v(x^+) \tau ^3 \; ,
\end{equation}
with $v$ independent of $x^-$.  As we shall see, in the constrained
case $v$ turns out to be independent of $x^+$ as well.

While one could simultaneously rotate $A^-\equiv A$ so that it has no
color 3 zero mode \cite{kap94,pik95}, we shall not do that here.
Instead we will retain this zero mode, which we call $w$, and
determine it in the solution of the equations of motion.

It is useful to introduce a set of transformations which are formally
the ``large'' gauge transformations which connect different Gribov
regions \cite{gri78}.  We shall denote these by $T^R_N$ and $T^L_N$,
with $N$ any integer:
\begin{eqnarray}
T^R_N &=& \exp\left[ -{ iN \pi \over 2L}  x^-\tau_3\right] \\
T^L_N &=& \exp\left[ { iN \pi \over 2L}  x ^+\tau_3\right]\; .
\end{eqnarray}
It is convenient to define the dimensionless variables $z_R= g v L/
\pi$ and $z_L= g w L/ \pi$, which $T^{R/L}_N$ shifts by $\pm N$:
\begin{eqnarray}
T^R_Nz_R (T^R_N)^{-1} &=& z_R + N
\label{whattdoesr} \\
T^L_Nz_L (T^L_N)^{-1} &=& z_L - N\; .
\label{whattdoes}
\end{eqnarray}
In addition, $T^{R/L}_N$ generates a space-time-dependent phase
rotation on the matter field $\phi _{R/L}$,
\begin{eqnarray}
T^R_N \phi _{R} (T^{R}_N)^{-1} &=& e ^{-iN\pi  x^-/ L} \phi _{R}
\label{tsymmr} \\
T^L_N \phi _{L} (T^{L}_N)^{-1} &=& e ^{ iN\pi  x^+/ L} \phi _{L}\; ,
\label{tsymm}
\end{eqnarray}
which however preserves the boundary conditions on $\phi_{R/L}$.  Note
that for $T_N^{R/L}$ to be a symmetry of the theory, the solutions for
$z_{R/L}$ in terms of the Fermi fields must correctly reproduce
(\ref{whattdoesr}) and (\ref{whattdoes}) under the transformations
(\ref{tsymmr}) and (\ref{tsymm}).  This will turn out to be the case.

The theory is also invariant under the so-called Weyl transformation,
denoted $R$.  This is also formally a gauge transformation, and takes
\begin{eqnarray}
Rz_{R/L}R^{-1} &=& -z_{R/L} \\
R \phi_{R/L} R^{-1} &=& \phi^\dagger_{R/L}\; .
\label{rsymm}
\end{eqnarray}

The action of $T^{R/L}_1$ and $R$ on the fermion Fock operators can be
determined easily from Eqns. (\ref{tsymmr}), (\ref{tsymm}) and
(\ref{rsymm}).  $T^R_1$ gives rise to a spectral flow for the
right-handed particles,
\begin{eqnarray}
T^R_1 b_n (T^R_1)^{-1} & = & b_{n-1} 
\qquad (n > 1/2) \nonumber \\
T^R_1 d_n (T^R_1)^{-1} & = & d _{n+1} \label{trsymm}\\
T^R_1 b _{{1\over2}} (T^R_1)^{-1} & = & d {_{{1\over2}} ^\dagger}
\; ,\nonumber
\end{eqnarray}
while $T^L_1$ gives rise to a spectral flow for the left-handed
particles,
\begin{eqnarray}
T^L_1 \delta_n (T^L_1)^{-1} & = & \delta _{n-1} 
\qquad (n > 1/2)\nonumber\\
T^L_1 \beta_n (T^L_1)^{-1} & = & \beta_{n+1} 
\label{tlsymm} \\ 
T^L_1 \delta_{{1\over2}} (T^L_1)^{-1} & = &
\beta{_{{1\over2}}^\dagger}\; .  \nonumber
\end{eqnarray}
The action of $R$ is analogous to charge conjugation,
\begin{eqnarray}
R b_n R^{-1} &=& - d_n  \label{fockrrh} \\
R \beta _n R^{-1} &=& - \delta _n\; .
\label{fockrlh}
\end{eqnarray}
The $a_N$ and $\alpha_N$ are invariant under both $T^{R/L}_1$ and
$R$. From the behavior of the Fock operators it is straightforward to
deduce the behavior of the elements of the Kac-Moody algebra under
$T^{R/L}_1$ and $R$, and show that the algebra is invariant.

It is also convenient to introduce a set of states $|V_M\rangle$,
where $M$ is any integer, which are related to one another by $T_N
\equiv T_N^R T_N^L$ transformations.  The Fock vacuum $|0\rangle$ is
defined to be $|V_0\rangle$, and
\begin{equation}
|V_M\rangle\equiv (T_1)^M|V_0\rangle\; ,
\end{equation}
with $(T_1)^{-1}=T_{-1}$.  It is straightforward to determine the
particle content of the $|V_M\rangle$ from the properties of the $T_1$
transformation \cite{pir96}.  One finds
\begin{eqnarray}
|V_N\rangle &=& d^\dagger_{N-\frac{1}{2}} 
\beta^\dagger_{N-\frac{1}{2}}\dots
d^\dagger_\frac{1}{2} \beta^\dagger_\frac{1}{2} |0\rangle
\label{nvacpos} \\
\nonumber\\
|V_{-N}\rangle &=& \delta^\dagger_{N-\frac{1}{2}} 
b^\dagger_{N-\frac{1}{2}} \dots
\delta^\dagger_\frac{1}{2} b^\dagger_\frac{1}{2} |0\rangle\; ,
\label{nvacneg}
\end{eqnarray}
for $N\geq0$.  These states are thus analogous to the ``$n$-vacua''
(\ref{vsubn})--(\ref{vsub-n}) found in the Schwinger model.

The operator $R$ satisfies $R^2=1$, so that its action on the Fock
vacuum may be defined to be
\begin{equation}
R|0\rangle = \pm|0\rangle\; .
\label{ronvac}
\end{equation}
Along with Eqns. (\ref{fockrrh}) and (\ref{fockrlh}), this choice
defines the action of $R$ on all states.  Without loss of generality
we may take the plus sign in Eqn. (\ref{ronvac}).  Then $R$
interchanges $|V_N\rangle$ and $|V_{-N}\rangle$:
\begin{equation}
R \vert V_N \rangle  =  (-1)^N \vert V_{-N} \rangle\; .
\label{rxchange}
\end{equation}
The factor $(-1)^N$ arises from the different ordering of the left-
and right-moving creation operators in (\ref{nvacpos}) and
(\ref{nvacneg}).

\section{Equations of Motion}

It is straightforward to derive the equations of motion for the theory
in the gauge we have chosen.  In the color helicity basis the Dirac
equation separates into
\begin{eqnarray}
\partial_-\psi_L &=& 0\\
\nonumber\\
\partial_-\phi_L &=& -igv\phi_L \label{eomphil}\\
\nonumber\\
\partial_+\psi_R &=& ig\left[A_-\phi_R - A_+\phi_R^\dagger\right]\\
\nonumber\\
\partial_+\phi_R &=& ig\left[A_+\psi_R - A_3\phi_R\right]\; .
\end{eqnarray}
In addition we have Gauss' law,
\begin{eqnarray}
-\partial {^2 _-} A_3 &=& gJ_3^R
\label{gauss3}\\
\nonumber\\
-(\partial _- \pm igv)^2 A_\pm &=& g J_\pm^R\; ,
\label{gausspm}
\end{eqnarray}
and Ampere's law,
\begin{eqnarray}
\partial_+ \partial_- A_3 
+ ig \left[ A_+(\partial_- - igv)A_-
        -A_-(\partial_- + igv)A_+ \right] &=& gJ^-_3
\label{ampere3}\\
\nonumber\\
\partial_+ \left[ (\partial_- \pm igv)A_\pm \right]
\pm ig\left[ A_3(\partial_-\pm igv)A_\pm
        -A_\pm \partial_-A_3 \right] &=& gJ^-_\pm\; .
\label{amperepm}
\end{eqnarray}
These latter relations will require particularly careful
consideration, as they explicitly connect left- and right-handed
quantities.  In particular we will find (as in the Schwinger model)
that some of these equations can only be satisfied in a subspace of
the full Hilbert space of the theory.  This subspace will be defined
to be the physical one.

The current operators appearing in (\ref{gauss3})--(\ref{amperepm})
include the gauge corrections computed in Eqn. (\ref{Ranomaly}).
Because of the gauge choice and the twisted boundary conditions,
however, the corrections to the color $\pm$ components of $J^R$
vanish.

We shall discuss the implementation of the Dirac equation and Ampere's
law below; for the moment let us consider the solution of
Eqns. (\ref{gauss3}) and (\ref{gausspm}).  As is usual in light-front
field theory, these relations mainly serve to determine the field $A$
on $x^+=0$ in terms of the dynamical degrees of freedom.  Note that in
the gauge we have chosen, Gauss' law separates and its individual
components can be solved directly; this is the motivation for
introducing the color helicity basis.

Eqn. (\ref{gauss3}) can be solved immediately to obtain the normal
mode part of $A_3$ on $x^+=0$:
\begin{equation}
A_3(0,x^-) = {gL \over 2\pi^2} \sum_{N =-\infty}^\infty 
\!\!\!\!{}^{\stackrel{\prime}{}}
\; {C^3_N \over N^2} e^{-iN \pi x^-/L}\; .
\end{equation}
The zero mode of $A_3$ is not determined by Eqn. (\ref{gauss3}).  We
shall return to this problem in a moment, but for now note that since
there is no zero mode on the left hand side of Eqn. (\ref{gauss3}),
the zero mode of the right hand side must also vanish.  This gives
\begin{equation}
0=C_0^3 - z_R \; .
\label{g3zm}
\end{equation}
This relation determines $z_R$ in terms of a charge operator, subject
to the same caveat we had in the Schwinger model: it is inconsistent
to take $z_R$ to be an operator involving right-moving fermions.  We
shall therefore rewrite Eqn. (\ref{g3zm}) in the form
\begin{equation}
0=C_0^3+D_0^3-D_0^3 - z_R \; ,
\end{equation}
and take
\begin{equation}
z_R = -D_0^3 \; .
\label{opsolnzr}
\end{equation}
What remains of the 3 color component of Gauss' law is then
\begin{equation}
C_0^3+D_0^3=0\; ,
\label{physsubs}
\end{equation}
which can be imposed as an eigenvalue condition defining physical
states.

Now let us consider Eqn. (\ref{gausspm}).  Inserting the operator
solution (\ref{opsolnzr}) for $z_R$ this becomes
\begin{equation}
-\left(\partial _- \mp {i\pi\over L} D_0^3\right)^2 A_\pm = g
J_\pm^R\; .
\label{gpmcon}
\end{equation}
The choice of twisted boundary conditions results in the covariant
derivatives having no zero eigenvalues, so they can be inverted to
give
\begin{equation}
A_\pm(0,x^-) = {gL \over 2\pi^2} \sum_{n=-\infty}^\infty 
{C^\mp_n \over \left (n \pm D_0^3 \right )^2}  e^{-in \pi x^-/L}\; .
\end{equation}
%

\section{Poincar\'e Generators}

As discussed previously, the Poincar\'e generators $P^\pm$ receive
contributions from both parts of the initial-value surface
[Eqn. (\ref{smppm})].  We shall denote the contributions coming from
integrating over $x^+=0$ and $x^-=0$ by $P^\pm_R$ and $P^\pm_L$,
respectively.

It is straightforward to work out the energy-momentum tensor following
the usual Noether procedure.  We obtain (with $\partial_-A^+=0$)
\begin{eqnarray}
\Theta^{+-} &=&-2 {\rm Tr} \left(F^{+-} \partial_+ A^+ \right )
-{\rm Tr}\left (F^{+-} F^{+-} \right)
+{i \over \sqrt{2}} {\rm Tr} \left( \Psi_R \partial^- \Psi_R -
\left(\partial^- \Psi_R\right) \Psi_R \right)\\
\Theta^{-+} &=&-2{\rm Tr}\left(F^{+-} \partial_- A^- \right )
-{\rm Tr}\left (F^{+-} F^{+-} \right)
+{i \over \sqrt{2}}{\rm Tr} \left( \Psi_L \partial^+ \Psi_L -
\left(\partial^+ \Psi_L\right) \Psi_L \right)\\
\Theta^{++} &=& {i \over \sqrt{2}}{\rm Tr} \left( \Psi_R \partial^+
 \Psi_R -\left(\partial^+ \Psi_R\right) \Psi_R \right)\\
\Theta^{--} &=& {i \over \sqrt{2}}{\rm Tr} \left( \Psi_L \partial^-
 \Psi_L -\left(\partial^- \Psi_L\right) \Psi_L \right)
-2{\rm Tr} \left(F^{-+} \partial_+ A^- \right )\;.
\end{eqnarray}
As in the Schwinger model, these lead to expressions we cannot
evaluate: they involve integrals of fields on surfaces where we do not
know them.  We will follow the rule discussed previously, and simply
drop the terms we do not know how to calculate.  In the end we shall
justify our results by showing that they correctly translate all
fields and satisfy the Poincar\'e algebra.

First let us construct $P^+$.  Using the rule of dropping terms we do
not know how to calculate, we obtain
\begin{equation}
P^+ = i\sqrt{2}\int_0^{2L} dx^- {\rm Tr} \left( \Psi_R \partial_-
\Psi_R \right)\; .
\label{p+}
\end{equation}
The operator product in this expression is singular and requires
regularization and renormalization.  This is accomplished as before by
splitting the product in $x^-$, introducing an appropriate exponential
factor to maintain gauge invariance.  We find
\begin{equation}
P^+ = \sum_{N>0}\left({N\pi \over L }\right) a{_N ^\dagger} a_N
+\sum_{n>0}\left({n\pi \over L }\right) \left[b{_n^\dagger} b_n + d{_n
^\dagger}d_n \right] - {\pi \over 2L} z_R^2\; ,
\label{qz}
\end{equation}
which leads to
\begin{equation}
P^+ = \sum_{N>0}\left({N\pi \over L }\right) a{_N ^\dagger} a_N
+\sum_{n>0}\left({n\pi \over L }\right) \left[b{_n^\dagger} b_n + d{_n
^\dagger}d_n \right] - {\pi \over 2L} (D_0^3)^2
\label{dynp+}
\end{equation}
when Eqn. (\ref{opsolnzr}) is used.  This expression will be tested
further for consistency below and found to be satisfactory.

Next let us discuss $P^-$.  The left-moving contribution is given by
\begin{equation}
P^-_L = {i \sqrt{2}}\int _0 ^{2L} dx^+ {\rm Tr} \left( \Psi_L
\partial_+ \Psi_L \right)\; .
\label{basicham}
\end{equation}
This operator product is singular, and is regulated by a
gauge-corrected splitting in $x^+$.  We find
\begin{equation}
P^-_L = \sum_{N>0}\left({N\pi \over L }\right) \alpha{_N ^\dagger}
\alpha_N +\sum_{n>0}\left({n\pi \over L }\right)
\left[\beta{_n^\dagger} \beta_n + \delta{_n ^\dagger}\delta_n \right]
-{g^2\over2\pi}\int_0^{2L}dx^+ {\rm Tr}(A^2)\; .
\end{equation}
Again, we do not know most of $A$ on the surface $x^-=0$ so we keep
only the part we do know on that surface.  This will turn out to be
the zero mode of $A_3$, which will be shown to be $x^+$-independent.
Thus we have
\begin{equation}
P^-_L = \sum_{N>0}\left({N\pi \over L }\right) \alpha{_N ^\dagger}
\alpha_N +\sum_{n>0}\left({n\pi \over L }\right)
\left[\beta{_n^\dagger} \beta_n + \delta{_n ^\dagger}\delta_n \right]
-{\pi\over2L}z_L^2\; .
\label{dynp-l}
\end{equation}
The contribution from the surface $x^+ = 0$ has the standard form one
expects in (1+1)-dimensional YM theory coupled to matter,
\begin{equation}
P^-_R = -g^2 \int ^{2L} _0 dx^- {\rm Tr} \left( J^+ {1\over D_-^2 }
J^+ \right)\; .
\end{equation}
None of these operator products are singular so evaluating this
expression is straightforward.  The result is most elegantly expressed
in terms of the $C$s.  We find
\begin{equation}
P^-_R = {g^2 L \over 4\pi^2} \left[\sum_{N =-\infty}^\infty
\!\!\!\!{}^{\stackrel{\prime}{}}
\thinspace\thinspace {1 \over N^2}C^3_N C^3_{-N} 
+ \sum_{n=-\infty}^\infty 
{1\over (n-D_0^3)^2}\left\{ C^+_n, C^-_{-n}\right\}\right]\; .
\label{dynp-r}
\end{equation}

There is a problem with this expression, however.  The presence of
$D_0^3$ in $P^-_R$ is in conflict with the (kinematical) Heisenberg
equation for $\phi_L$.  We should have
\begin{equation}
-i[\phi_L(x^+,0),P^-] = \partial_+\phi_L(x^+,0)\; ,
\label{philheis}
\end{equation}
which is already accomplished by the free part of $P^-_L$.  Since
$[\phi_L,D_0^3] = \phi_L$ the interaction terms in $P^-_R$ spoil
(\ref{philheis}).  To cure this problem we can simply modify $P^-_R$
by replacing $D_0^3$ with an operator that is equal to it in the
physical subspace.  The natural thing to try is the substitution
$D_0^3\rightarrow -C_0^3$, motivated by Eqn. (\ref{physsubs}).  This
is potentially in conflict with the Dirac equation for $\phi_R$,
however.  Before the substitution we obtained the correct commutator
of $\phi_R$ with $P^-$, but afterwards the fact that $[\phi_R,C_0^3]
\neq 0$ leads to a new and unwanted term in $[\phi_R,P^-]$:
\begin{equation}
[\phi_R,P^-] = {g^2L\over 4\pi^2}\sum_n \left[\phi_R,{1\over
\left(n+C_0^3\right)^2}\right] \left\{C^+_n,C^-_{-n}\right\}+\dots\; ,
\end{equation}
where the dots represent the terms we had originally.  Now this extra
term can be shown to vanish in matrix elements between states
satisfying (\ref{physsubs}), but this happens for a fairly trivial
reason: it is a colored operator and the physical states are required
to be colorless.  A less trivial check is to consider the commutator
of a colorless operator such as $\phi_R^\dagger\phi_R$ with our new
$P^-$.  In fact the commutator of this operator also reduces, in the
subspace defined by (\ref{physsubs}), to what we obtain from the Dirac
equation, so the modified $P^-$ appears to be consistent.  Presumably
one wants all relations derivable from the equations of motion to be
recovered in the physical subspace.  It would be helpful to have a
more detailed understanding of this point, as well as the extent to
which (\ref{dynp-r}) satisfies the necessary conditions.

Eqns. (\ref{dynp+}), (\ref{dynp-l}) and (\ref{dynp-r}), with
$D_0^3\rightarrow -C_0^3$, are our trial forms for $P^\pm$.  Our next
task is to check whether these correctly reproduce the Dirac equation
and Ampere's law for this theory.  This is a straightforward exercise
in commuting fields with $P^\pm$ and comparing the results with the
corresponding equations of motion.

It turns out that the Dirac equation for $\psi_R$ is satisfied if
\begin{equation}
z_L = -C^3_0\; ,
\label{cl}
\end{equation}
which we shall take to be a strong (operator) equality.  This
determines the zero mode of $A_3$, which we were not able to fix using
Gauss' law.  Note that $z_L$ commutes with $P^-$, and so is
$x^+$-independent as promised.  

Next, Ampere's law is satisfied if
\begin{equation}
D_N^3 = D_n^\pm = 0\; .
\end{equation}
These conditions must be realized weakly, in matrix elements between
states.  We shall require physical states to satisfy
\begin{equation}
D_N^3 |\Phi\rangle = 0\qquad (N>0)\; ,
\label{dn3+-}
\end{equation}
in analogy with the Schwinger model.  Note, however, that due to
\begin{equation}
[(C_0^3+D_0^3),D_n^\pm] = \pm D_n^\pm
\end{equation}
(see Eqn. (\ref{kacmoody3pm})), matrix elements of $D_n^\pm$ between
states that satisfy (\ref{physsubs}) are automatically zero.  It is
therefore not necessary to impose the condition $D_n^\pm\approx 0$
separately.

Finally, the zero mode of the color 3 component of Ampere's law
reduces to
\begin{equation}
{\cal P}\equiv\sum_n {1 \over \left (C_0^3+n \right)^3} \left \{C_n^+,
C_{-n}^-\right \}=0\; ,
\label{finalcond}
\end{equation}
which again must be realized in matrix elements between physical
states.  To see that Eqn. (\ref{finalcond}) is in fact satisfied in
the physical subspace, let us discuss the physical states in more
detail.  These will be obtained by acting with gauge-invariant
operators built from the right-handed fields on the physical vacuum
state $|\Omega\rangle$:\footnote{The conditions (\ref{dn3+-}) remove
from the physical subspace any states with a left-handed particle
content beyond what is present in the vacuum.}
\begin{equation}
|\Phi\rangle = {\cal O}|\Omega\rangle\; .
\end{equation}
Now, the transformation $R$ is a symmetry of the theory, as is easily
verified from the expressions for $P^\pm$.  Without loss of generality
we may choose the vacuum to be an eigenstate of $R$ with eigenvalue
$+1$.  Since a gauge-invariant operator is in particular invariant
under $R$,
\begin{equation}
R{\cal O}R = {\cal O}\; ,
\end{equation}
it follows that all physical states are also eigenstates of $R$ with
eigenvalue $+1$.  But ${\cal P}$ is odd under $R$,
\begin{equation}
R {\cal P}R = -{\cal P} \; ,
\end{equation}
and therefore matrix elements of ${\cal P}$ between physical states
are zero as required.

The final result is that it appears to be consistent to take $P^+$ as
given in Eqn. (\ref{dynp+}) and
\begin{eqnarray}
P^- &=& \sum_{N>0}\left({N\pi \over L }\right) \alpha{_N ^\dagger}
\alpha_N +\sum_{n>0}\left({n\pi \over L }\right)
\left[\beta{_n^\dagger} \beta_n + \delta{_n ^\dagger}\delta_n \right]
-{\pi\over2L}(C_0^3)^2\nonumber\\ & &+ {g^2 L \over 4\pi^2} \left[
\sum_{N =-\infty}^\infty \!\!\!\!{}^{\stackrel{\prime}{}}
\thinspace\thinspace {1\over N^2}C^3_N C^3_{-N}
+\sum_{n=-\infty}^\infty {1\over (n+C_0^3)^2} 
\{C^+_n, C^-_{-n}\}\right]\; .
\end{eqnarray}
Physical states have vanishing color 3 charge,
\begin{equation}
(C^3_0+D^3_0)|\Phi\rangle = 0\; ,
\end{equation}
and satisfy Eqn. (\ref{dn3+-}).  One can check that the operators that
annihilate physical states commute with $P^+$ and $P^-$, so the
physical subspace is stable.  It can also be shown that $[P^+,P^-]=0$,
as required.

\section{Axial Anomaly}

As a further check of the formulation, let us now discuss the axial
anomaly in this theory.  We shall focus on the color 3 part of the
currents, for which the anomaly relation reads
\begin{equation}
\partial _{\mu} J^{\mu,3}_5 = {g \over 2 \pi} \epsilon ^{\mu \nu}
F_{\mu \nu}^3 \; .
\label{anomalyreln}
\end{equation}
In 1+1 dimensions, the axial current $J^{\mu}_5 = -[\Psi,
\gamma^\mu\gamma^5\Psi]/\sqrt{2}$ is related to the vector current
$J^\mu$ through $J^{\mu}_5 = (J^{R}, -J^L)$. In addition, it can be
shown that covariant derivatives reduce to partial derivatives for
$J^{\mu,3}$ so the conservation equations below take the Abelian form
for the matter currents.

To check Eqn. (\ref{anomalyreln}) we first calculate
$\partial_+J{^{R}_3}$ from
\begin{equation}
i[P ^- ,{J}^{R}_3(x)] = \partial _+ {J}^{R}_3 (x)\; .
\end{equation}
Using
\begin{equation}
[\tilde {J} {^{R} _3} (0,x^-) , \tilde {J} {^{R} _3} (0,y^-)] = {i
\over 2 \pi} \delta'(x^- - y^-)
\end{equation}
we find
\begin{equation}
\partial_+ {J} {^{R} _3}(0,x^-) ={g \over 2 \pi} \partial _- A
_3(0,x^-) \; ,
\end{equation}
where we have chosen to evaluate the currents at $(0,x^-)$.
Similarly, we can compute $\partial_-J^L_3$ by commuting it with
$P^+$.  Since $\tilde{J}^L _3 $ is independent of $x^-$ by the
equations of motion, the only contribution comes from the gauge
correction to $J^L_3$.  Thus we find
\begin{equation}
\partial _- J{_3 ^L}(0,x^-) = -{g \over 2 \pi} \partial _- A_3(0,x^-)
\; .
\end{equation}
Combining these results we then obtain
\begin{eqnarray}
\partial _{\mu} J {_ 3 ^{\mu}} &=& \partial _+ J {_3 ^{R} } +\partial _-
J{_3 ^L} \nonumber\\
&=& 0
\end{eqnarray}
and
\begin{eqnarray}
\partial _{\mu} J {^{5, \mu} _3} &=& \partial _+ J {_3 ^{R}} -
\partial_- J {_3^L} = {g \over \pi}\partial _- A_3
\nonumber\\
&=& {g \over 2 \pi} \epsilon ^{\mu \nu} F_{\mu \nu}^3\; ,
\end{eqnarray}
as expected.  The formulation of the theory therefore appears to be
satisfactory, and we can now study the structure of the ground state.

\section{Vacuum States}

It is straightforward to verify that the states $|V_N\rangle$ are all
degenerate (they have $P^+=P^-=0$) and lie in the physical subspace.
The physical vacuum states will thus be appropriate superpositions of
these, constructed to be (phase) invariant under the residual $T_1$
and $R$ symmetries.

The most general superposition consistent with $T_1$ invariance is
simply
\begin{equation}
|\theta\rangle = \sum_{N=-\infty}^\infty e^{-iN\theta} |V_N\rangle\; ,
\label{constrvac}
\end{equation}
which satisfies
\begin{equation}
T_1|\theta\rangle = e^{i\theta}|\theta\rangle\; .
\end{equation}
Acting with the Weyl transformation $R$ then gives
\begin{eqnarray}
R|\theta\rangle &=& \sum_{N=-\infty}^\infty e^{-iN\theta}
(-1)^N|V_{-N}\rangle\nonumber\\
&=& \sum_{N=-\infty}^\infty e^{iN(\theta-\pi)} |V_N\rangle\; .
\end{eqnarray}
This is equal to $|\theta\rangle$ up to a phase only for $\theta =
\pm\pi/2$.  We therefore have a pair of distinct physical vacuum
states labeled by a discrete vacuum angle.  We shall refer to these
physical vacua as $|\Omega_\pm\rangle$.

We have here motivated the formation of the superposition
(\ref{constrvac}) as a way of resolving the residual (large) gauge
invariance of the theory.  It is presumably also necessary to build
the theory on a vacuum of this form in order to satisfy the cluster
property, as in the Schwinger model.  To verify this explicitly for
the present model, however, we would need to do a more complete
dynamical calculation.

Let us now consider whether this vacuum structure has any affect on
the spectrum of the theory, that is, whether the spectrum depends on
the vacuum angle $\theta=\pm\pi/2$.  Consider calculating a matrix
element of $P^-$ between any two physical states:
\begin{equation}
\langle\Omega_\pm|{\cal O}^\prime (P^-_R+P^-_L) {\cal O}
|\Omega_\pm\rangle\; ,
\end{equation}
where ${\cal O}$ and ${\cal O}^\prime$ are gauge-invariant operators
constructed from the right-handed fields.  Since $P_L^-$ commutes with
these it simply passes through to act on the vacuum, where it gives
zero.  In addition, since ${\cal O}$ and ${\cal O}^\prime$ contain no
left-handed fields, the left-handed particles in the vacuum serve to
``diagonalize'' the matrix element between the different
$|V_M\rangle$:
\begin{equation}
\langle\Omega_\pm |{\cal O}^\prime P^-_R {\cal O}|\Omega_\pm\rangle =
\sum_{N=-\infty}^\infty \langle V_N| {\cal O}^\prime P^-_R {\cal O}
|V_N\rangle\; .
\label{melofp-}
\end{equation}
Now ${\cal O}$ and ${\cal O}^\prime$ are invariant under $T_1$, and
furthermore it can be shown that
\begin{equation}
T_1 P^-_R T_1^{-1} = P^-_R\; .
\end{equation}
Inserting factors of $(T_1^{-1} T_1)^N$ between the states and
operators we therefore find
\begin{equation}
\langle V_N|{\cal O}^\prime P^-_R {\cal O}|V_N\rangle =
\langle V_0|{\cal O}^\prime P^-_R {\cal O}|V_0\rangle\; .
\end{equation}
All the matrix elements on the right hand side of (\ref{melofp-}) are
thus identical, and so
\begin{equation}
\langle\Omega_\pm|{\cal O}^\prime P^-_R {\cal O}|\Omega_\pm \rangle =
\langle V_0|{\cal O}^\prime P^-_R {\cal O}|V_0\rangle\; ,
\end{equation}
up to the (infinite) normalization factor necessary for the state
$|\Omega_\pm\rangle$.  Finally, we note that $C_0^3|V_0\rangle=0$, so
that we may replace $C_0^3$ by zero in the expression for $P^-_R$.
The resulting $P^-_R$ is simply the usual DLCQ Hamiltonian for this
theory, that is, with the zero modes discarded.  The final result is
that matrix elements of the Hamiltonian in the full theory are equal
to those we would obtain by taking the trivial Fock vacuum
$|V_0\rangle$ and ignoring the zero modes.  Thus the standard DLCQ
procedure gives the correct spectrum for this theory.

We should perhaps emphasize that in more complicated theories, such as
QCD, the analogous result will certainly not hold.  In that case there
are physical quantities that do depend on the vacuum angle $\theta$,
and it will not be possible to correctly reproduce these by neglecting
the vacuum structure.

\section{The Condensate}

It is generally believed that YM$_{1+1}$ coupled to adjoint fermions
develops a condensate $\Sigma \equiv\langle\Omega | {\rm Tr}
(\bar{\Psi}\Psi) |\Omega \rangle$.  Thus far $\Sigma$ has been
calculated in the large-$N$ limit \cite{koz95} and in the small-volume
limit for SU(2) using equal-time quantization \cite{les94}.  A
condensate was also computed in a chiral version of this theory
containing only right-handed fermions \cite{pir96}.  In that
calculation it was the field itself which acquired a vacuum
expectation value and the result was fundamentally different from what
we are considering here.

It is straightforward to compute $\Sigma$ in the vacuum
(\ref{constrvac}).  To be specific, we shall evaluate $\Sigma$ at the
space-time point $(0,x^-)$.  The terms in ${\rm Tr} (\bar{\Psi}\Psi)$
that can contribute to the VEV are
\begin{equation}
{\rm Tr}(\bar{\Psi}\Psi) = i\left(\phi_L^\dagger\phi_R -
\phi_R^\dagger\phi_L\right) + \cdots\; .
\end{equation}
These operator products are not singular, so point splitting is not
required.  

The field $\phi_R(0,x^-)$ is just given by its initial value
(\ref{phirinitial}), of course.  To obtain $\phi_L$ at $(0,x^-)$ we
must solve its equation of motion, i.e., Eqn. (\ref{eomphil}), with
the noncommuting factors on the RHS symmetrized:
\begin{equation}
\partial_-\phi_L = {i\pi\over 2L}\left(D_0^3\phi_L +
\phi_LD_0^3\right)\; .
\end{equation}
Integrating this gives $\phi_L(0,x^-)$ in terms of its initial value
at the corner point $(0,0)$:
\begin{equation}
\phi_L(0,x^-) = e^{i\pi x^- D_0^3/2L} \phi_L(0,0) 
e^{i\pi x^-D_0^3/2L}\; .
\label{philsolved}
\end{equation}
With these results it is straightforward to evaluate the condensate;
we find
\begin{equation}
\langle\Omega_\pm | {\rm Tr} (\bar{\Psi}\Psi) |\Omega_\pm \rangle =
\pm {1\over \sqrt{2}L}\; .
\end{equation}
Note that the result is independent of $x^-$, as it should be.  This
happens because the exponential factors in (\ref{philsolved}) acting
on the $|V_N\rangle$ exactly compensate the exponentials from the
field expansion (\ref{phirinitial}).\footnote{Note that use of the
properly symmetrized solution (\ref{philsolved}) is crucial for this
to work!}  In addition, however, the condensate is proportional to
$1/L$, and so vanishes in the continuum limit.

This type of behavior is also found in the light-front version of the
Schwinger model discussed in Refs. \cite{mcc91}.  In that case the
$1/L$ behavior can be traced to the crude treatment of the small-$p^+$
region---in particular the violation of parity---that results when
periodicity conditions are imposed on null planes \cite{mcc96}.  In
the Schwinger model the problem can be cured by forcibly
parity-symmetrizing the theory after it has been solved.  In the
present example this sort of approach may not be practical, since the
model is not analytically soluble.

As discussed previously, the condensate, or more precisely the
presence of nontrivial vacuum structure, has no effect on the mass
spectrum of the theory.  This result is in accordance with recent work
of Kutasov and Schwimmer \cite{kus95}, who claim that there are
classes of two-dimensional YM theories which have the same massive
spectrum.  A necessary condition for this universality is the
decoupling of the massless (vacuum) and massive sectors.  Our
construction exhibits this directly.  In particular, the massive
spectrum may be obtained by neglecting the vacuum structure and gauge
field zero modes, that is, by applying the naive light-front
formalism.  In addition, the only left-moving quanta that enter
physical states reside in the vacuum; the physical excitations are
built entirely from the right-movers.

It would be interesting to study whether the spectrum depends on the
vacuum angle when a fermion mass is turned on, as for example occurs
in the massive Schwinger model.  At present we have nothing definite
to say on this question, although the condensate $\Sigma$ does arise
in at least one interesting context in the massive theory.  It has
been shown recently that two-dimensional gauge theories with massless
fermions can screen ``fractional'' test charges---charges in
representations of the gauge group that are smaller than the one
carried by the dynamical fermions in the theory \cite{grk95}.  Thus,
for example, in the theory with adjoint fermions the fundamental
Wilson loop exhibits a perimeter-law behavior.  When the dynamical
fermions are given a mass $m$, however, the screening disappears and
fractional charges are confined, with a string tension
\begin{equation}
\sigma =2 m \Sigma\; .
\end{equation}
It is unclear whether this has implications for the question of
whether the spectrum of the theory with massive fermions itself is
affected by the condensate \cite{grk95}.

\section{Conclusions}

We have shown that the $Z_2$ vacuum structure of SU(2) gauge theory
coupled to adjoint fermions in 1+1 dimensions can in fact be
successfully reproduced in the light-front framework.  We have found a
pair of (degenerate) physical vacuum states, and a nonzero ``chiral''
condensate which is sensitive to the vacuum physics.  This vacuum
structure decouples from the massive spectrum, however, consistent
with Ref. \cite{kus95}.

Three ingredients are essential for obtaining this structure.  First,
it is necessary to include a complete set of independent degrees of
freedom, that is, independent fields initialized on two different null
planes.  The surface $x^+=0$ does not define a Cauchy problem for the
left-moving degrees of freedom.  For some purposes it may be possible
to ignore this subtlety.  In the present case, for example, the
physics of massive states can be correctly recovered by ignoring the
left-movers and zero modes.  This is a feature peculiar to
(1+1)-dimensional gauge theories with massless matter, however
\cite{kus95}.  It is not expected to be true in more complicated
theories like QCD, where there is a strong coupling between massless
and massive states.

Second, it is important to pay close attention to the interplay
between the gauge choice and boundary conditions.  Given the
periodicity conditions we have imposed to regulate the theory, it is
not possible to gauge all the zero modes of $A_\mu$ to zero.  Certain
of these must be retained in the theory and their properties
determined.  This situation is quite familiar even in equal-time
quantization, when one regulates with equal-time periodicity
conditions and attempts to impose a spacelike axial gauge (see, e.g.,
\cite{nak83,man85}).

Finally, it is necessary to carefully define singular operator
products in a gauge-invariant way.  The resulting gauge corrections to
$P^\pm$ are what allow certain states which contain pairs of right-
and left-moving quanta to actually be degenerate with the bare vacuum.
Furthermore, the gauge corrections to the current operators are
crucial for determining the gauge field zero modes, as well as for
obtaining the correct anomaly relation.

The vacuum states have a much simpler structure in the light-front
representation than at equal time.  (This is also true in the
Schwinger model \cite{mcc91}.)  However, the formulation with
periodicity conditions on the characteristic surfaces has the property
that some of the details of the operator products, such as the
condensate, do not approach their continuum values as the periodicity
length is taken to infinity.  The degree to which one may lose the
ability to represent some aspects of the physics in the discretized
light-front approach is not entirely understood.  It would be very
interesting to have a more general understanding of this point.

It would also be of interest to extend this sort of construction to
the case of massive fermions, where the vacuum structure can play a
more meaningful physical role.  In the massive Schwinger model, for
example, the spectrum of states depends on the vacuum angle $\theta$,
a property shared by QCD.

Finally, it will not have escaped the reader that the construction we
have presented relies to an uncomfortable degree on trial-and-error.
We know of no standard procedure, analogous to the textbook treatment
for equal-time field theory, which leads directly to the correct
dynamical operators.  Instead, we start from the canonical expressions
for $P^\pm$, discarding terms we do not know how to evaluate and
including gauge corrections arising from the renormalization of
singular operator products.  Further modifications may be necessary in
response to checks of consistency, in particular the replacement of
troublesome operators with operators to which they are weakly
equivalent, i.e., equal in the physical subspace.  Of course, the
definition of the physical subspace, and hence which operators may be
considered to be weakly equivalent, itself depends on the form of the
Poincar\'e generators: these determine which Heisenberg equations are
not obtained directly as operator relations, and thereby fix the
conditions which must be satisfied by states in order for these to
hold in a weak sense.  The procedure thus has an unpleasantly circular
character.  It would be of great interest to have a more
straightforward formulation of light-front field theory, particularly
as the addition of further space-time dimensions, and the associated
renormalization problems, can only increase the difficulties.

\acknowledgments
\noindent
It is a pleasure to thank David Kutasov, Pierre van Baal and Ariel
Zhitnitsky for helpful discussions.  S.S.P. would also like to
acknowledge the hospitality of the Aspen Center for Physics, where
part of this work was completed.  This work was supported in part by
grants from the U.S. Department of Energy.

\end{document}